\documentclass[sigconf]{acmart}

\usepackage{booktabs} 

\renewcommand\footnotetextcopyrightpermission[1]{} 
 \setcopyright{none}
\begin{document}
\title{Stroke-based sketched symbol reconstruction and segmentation}

\author{Kurmanbek Kaiyrbekov}
\orcid{1234-5678-9012}
\affiliation{%
  \institution{Johns Hopkins University}
  \city{Baltimore} 
  \country{MD, USA} 
  \postcode{21218}
}
\email{kurmanbek@jhu.edu}

\author{T. Metin Sezgin}
\affiliation{%
  \institution{Koc University}
  \streetaddress{College of Engineering, Koc University}
  \city{Istanbul} 
  \country{Turkey} 
  \postcode{34450}
}
\email{mtsezgin@ku.edu.tr}

\renewcommand{\shortauthors}{K. Kaiyrbekov and T. M. Sezgin}

\begin{abstract}
Hand-drawn objects usually consist of multiple semantically meaningful parts. For example, a stick figure consists of a head, a torso, and pairs of legs and arms. Efficient and accurate identification of these subparts promises to significantly improve algorithms for stylization, deformation, morphing and animation of 2D drawings. In this paper, we propose a neural network model that segments symbols into stroke-level components. Our segmentation framework has two main elements: a fixed feature extractor and a Multilayer Perceptron (MLP) network that identifies a component based on the feature. As the feature extractor we utilize an encoder of a \textsf{stroke-rnn}, which is our newly proposed generative Variational Auto-Encoder (VAE) model that reconstructs symbols on a stroke by stroke basis. Experiments show that a single encoder could be reused for segmenting multiple categories of sketched symbols with negligible effects on segmentation accuracies. Our segmentation scores surpass existing methodologies on an available small state of the art dataset. Moreover, extensive evaluations on our newly annotated big dataset demonstrate that our framework obtains significantly better accuracies as compared to baseline models. We release the dataset to the community.
\end{abstract}

%
%

\ccsdesc[500]{Computing methodologies~Object Identification}
\ccsdesc[300]{Computing methodologies~Sketch recognition}
\ccsdesc[100]{Human-centered computing~Human computer interaction (HCI)}

\keywords{Sketches, Segmentation, Neural Networks}

\maketitle
\section{Introduction}

From ancient times, sketching has been used by humans as a natural channel of communication. Current ubiquity of touchscreen devices provided an accessible and contemporary medium for sketching, thus there has been a growing research interest in sketch-based applications. There are three important research directions that have plethora of potential applications: categorizing sketches, drawing sketches, and segmenting sketches into semantically meaningful components. Most of the previous work focused on the categorization problem of determining object class of an input drawing \cite{sezgin2005hmm,tirkaz2012sketched,schneider2014sketch}. This is a challenging task even for humans due to varying levels of abstraction of drawings, lack of visual cues and similarity of different categories of sketched symbols. However, utilization of discriminative Deep Neural Network (DNN) models allowed machines to surpass human performance on the categorization problem \cite{Yu:2017:SDN:3086064.3086091,seddati2015deepsketch}.  In this article, we utilize powerful modeling capability of generative neural networks to segment symbols into stroke-level components. 

The generative models are essential and active area of deep learning research. DNN models like Generative Adversarial Network (GAN) \cite{NIPS2014_5423}, Deep Recurrent Attentive Writer (DRAW) \cite{gregor2015draw}, and VAE \cite{kingma2013auto} made dramatic improvements in generative modeling of pixel images. Images consist of millions of pixels which are captured all at once by a sophisticated cameras. Sketches represented as vector images, on the other hand, are considerably smaller in size and they consist of single or several sequentially drawn strokes. Therefore, it is computationally more efficient to use the vector representation while processing the drawings. 

The release of Google, QuickDraw! dataset fostered application of the generative models on vector images. Along with dataset Ha and Eck \cite{sketchrnn} have introduced a sketch-rnn, it is a sequence-to-sequence VAE framework that encodes vector representation of the sketches using bidirectional recurrent neural network (Bi-RNN) \cite{schuster1997bidirectional} and reconstructs (or decodes) with another RNN network. More recently, Chen et al. \cite{chen2017sketch} proposed similar VAE model which replaced the Bi-RNN encoder with a Convolutional Neural Network (CNN) and removed Kullback-Lieber divergence term from the objective function of the VAE. Both of these models take a whole sketch as an input either in vector image or pixel image format and reconstruct vector representation of a drawing. These two models are able to draw categories that they were trained on and their performance deteriorates if an input sketch is significantly different in appearance. However, sketches are composed of multiple sequential strokes and people can use the same set of strokes to draw disparate categories of sketches. For example, if a person knows how to draw a circle he/she should be able to sketch eyes of a cat, wheels of a car, whole pizza etc. Motivated by this, we propose a stroke level VAE model \textsf{stroke-rnn} which learns to reconstruct symbol strokes. Hence, it is capable of drawing many categories even if trained only on a single object category.

Segmenting a symbol into semantically meaningful components is hard because identical strokes could represent distinct subparts. For instance, a stroke in a shape of circle could represent a head of a cat, an eye or even a mouth. Information about contour and location of the stroke combined could provide necessary cues for identifying a component correctly. The \textsf{stroke-rnn} can encode the stroke into a vector from which the stroke is reconstructed at an appropriate position with a proper shape, so the encoded vector carries a sufficient information about the location and shape. We propose a neural network model that takes the output of the encoder of the \textsf{stroke-rnn} as an input feature, and then identifies the component the stroke belongs to. People use a single stroke to draw multiple components, a full component or part of a component. Our focus is on latter two cases, though the case of multiple components could be handled by segmenting stroke into components either automatically or manually. Due to lack of a comprehensive dataset of segmented sketches, previously, researchers tested their frameworks on small number of labeled symbols that are far more complex than the ones collected by Google, Quick, Draw!. Therefore, for proper evaluation, we annotate 500 sketches for 5 categories. We show success of our framework both on our labeled dataset compared to our newly proposed baseline model and on a previously used small but complex dataset by Huang et al. \cite{huang2014data} compared to scores reported by previous works. In summary, the main contributions of this paper are as follows:

\begin{itemize}
\item Generative neural network model that can reconstruct multiple disparate object classes of sketches.
\item A new big dataset of sketches with annotated semantically meaningful components.
\item A baseline model for segmentation of vector representation of drawings.
\item The neural network framework for symbol segmentation that outperforms baseline model for our dataset and previous methods on the old dataset.
\end{itemize}

The rest of the paper is organized as follows: First we introduce previous generative reconstruction and segmentation models for sketches. Then in Section \ref{Reconstruction Model} we describe our stroke based reconstruction model and in Section \ref{Experiments on Reconstruction Model} we demonstrate the performance of the model using several experiments. In Section \ref{Segmentation Model} we describe our and neural network frameworks for symbol segmentation. We show the results of the experiments performed on the segmentation models in Section \ref{Experiments on Segmentation}. Finally, we conclude with the summary and discussion of future work. 
\begin{figure*}
\includegraphics[width=\textwidth,height=\textheight,keepaspectratio]{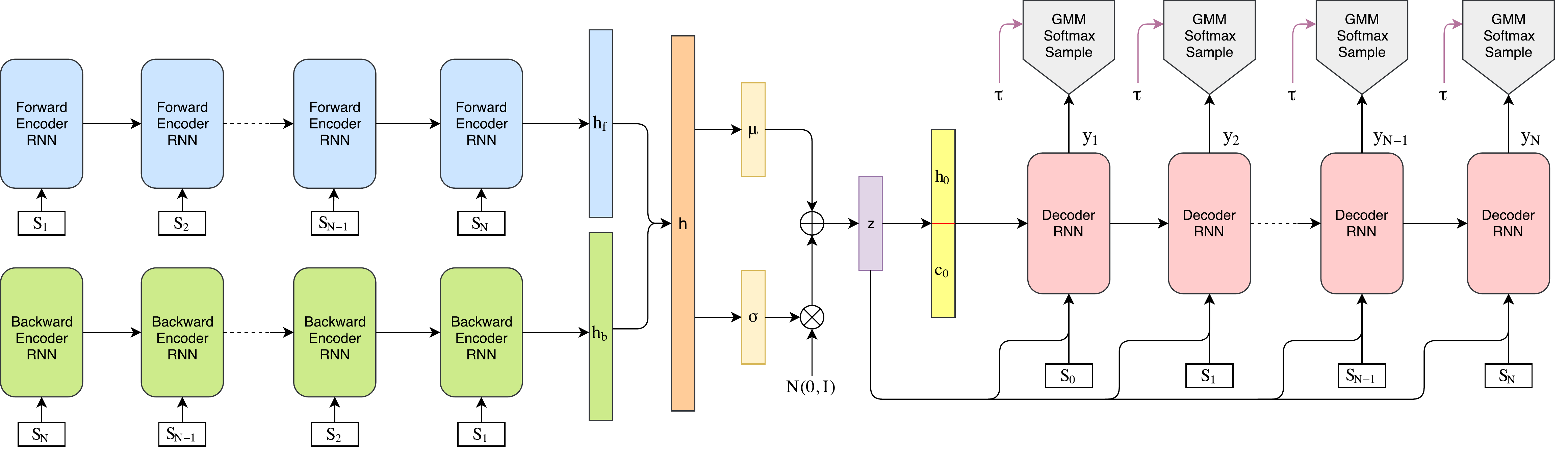}\caption{Architecture of the stroke-rnn.}
\label{strokernn}
\end{figure*}

\section{Related work} \label{Related Work}
The discussion of the related work is divided into two parts. First, we review the relevant work on sketch generation. Then, we discuss previous work associated with sketch segmentation.
\subsection{Reconstruction by generative models}
Neural network research community mostly focused on generative modeling of images or sentences. Generative adversarial frameworks like LAPGAN \cite{denton2015deep} and DCGAN \cite{radford2015unsupervised} have shown to be able to generate high-quality images. VAE is another widely used framework was shown to successfully to generate image captions \cite{pu2016variational} sentences  \cite{bowman2015generating} and sentence translations \cite{zhang2016variational}. 

Pioneering work on generative modeling of sketches was performed by Ha and Eck who, along with a release of the large Google, Quickdraw! dataset, have proposed the first generative framework based on VAE. Their model encoded and decoded vector representation of sketches. This work was followed by VAE-based sketch-pix2seq framework where Chen et. al. made two important modifications to sketch-rnn. First, they replaced a Bi-RNN encoder with CNN-based one and encoded 48x48 monochrome png images of sketches instead of vector representations. Second, they removed KL-divergence term from objective function. Based on the results of Turing tests they argued that sketch-pix2seq is capable of learning and generating multiple categories of sketches better than sketch-rnn. In contrast our \textsf{stroke-rnn} VAE-based model learns to encode and reconstruct lower level strokes of vector representation of symbols. Since drawings are composed of multiple strokes of shorter lengths, our method is a better suit for recurrent neural network training because RNNs suffer from vanishing or exploding gradients for long sequences \cite{bengio1994learning}. Moreover, since multiple disparate categories could be drawn using same set of strokes our model is able to reconstruct numerous object categories even if was not trained on them.

\subsection{Segmentation of sketched symbols}  
Most of the previous work on semantic understanding of sketches focused on object level classification task, which is a task of finding distinct objects given a sketched scene with several objects, rather than segmentation of particular sketches. For instance, Sun et al. \cite{sun2012free} published an article that focused on object level segmentation of free hand sketches. They proposed a graph-based sketch segmentation algorithm which made use of a million clip art images collected from the web as a knowledge database. They used greedy segment merging strategy to extract sketch objects from scenes. On the other hand, we focus on extracting meaningful subparts from a known sketch object. 

Achievements of assembly-based 3D modeling in segmentation and labeling of images have motivated Huang et al. to successfully apply these methods on the problem of sketch segmentation. They designed a part-assembly approach for sketch interpretation that matched segments of the sketch to corresponding components on 3D meshes from image database. Despite success of their method, its main drawback was that the model needed a database of readily segmented and labeled corresponding images, which makes it difficult to enlarge their sketch dataset. Moreover, their best model relied on human assistance to align the sketch to its corresponding 3D mesh. Then, Schneider and Tuytelaars \cite{schneider2016example} proposed a segmentation framework based on graphical models. They developed a heuristic to encode relations between strokes as a graph and used Conditional Random Fields (CRF) \cite{lafferty2001conditional} construct the most probable part level segmentation of a sketch. First, they segmented strokes of the sketches at high curvature points. Then, they used Fisher Vectors \cite{sanchez2013image} with a concatenated spatial information of a stroke as a feature vector and Support Vector Machines (SVM) \cite{cortes1995support} to get classification probabilities for the stroke. They applied Platt's scaling \cite{platt1999probabilistic} to convert output score of SVM into probabilistic output. Finally, they constructed graph and used CRFs to find an overall most probable component level configuration of the sketch. Recently, Li et al. \cite{LLi2018} have used CNN with subsequent refinement based on graph cuts to segment drawings on the Huang et al. dataset. They also treated symbols as a pixel images and demonstrated superiority of neural network model over standard approaches. All of the aforementioned models treat drawings as pixel images. However, we use computationally advantageous vector representation of symbols. Despite being simple and flexible our model demonstrates better accuracies than previous methods, which makes it a better suit for possible applications. 

Even more recently, Wu et al. \cite{SketchSegNet} suggested a stroke-level segmentation model for symbols. They labeled  60 random sketches for seven categories of QuickDraw! datset, augmented their dataset using sketch-rnn, then used VAE network to get a component label for each point. The overall label of the stroke was decided based on amount of correctly labeled points. Our network on the other hand labels whole stroke. It uses a single encoder as a fixed feature extractor and needs to train only a simple 3 layer segmentation MLP network. Nevertheless, our framework obtains state of the art accuracies on previously labeled dataset by Huang et al.. We thoroughly evaluate our model on a new comprehensive dataset that contains 500 annotated symbols per category. Our MLP-baed framework obtains much better scores as compared to the best baseline methods on our dataset.

\section{Generative Reconstruction Model} \label{Reconstruction Model}

\subsection{Dataset}
We use simplified drawings from the dataset collected using Google, Quick, Draw!, an online game in which users were asked to draw a specific object in 20 seconds. Symbols consist of ordered set of strokes, while strokes made up of ordered sequence of points. Unlike previous methods, we train our model on individual strokes of particular symbol category instead of full symbol sequences. As Ha and Eck, we represent each point in a stroke via a 5-D vector $[ \Delta x \ \Delta \ y\ p_1 \ p_2 \ p_3 ]$ where $ \Delta x  = x_{cur} - x_{prev}, \ and \ \Delta \ y = y_{cur} - y_{prev} $ are displacements of current point with respect to the previous point. Each of the pen states $ p_1 \ p_2 \ and \ p_ 3$ are binary variables. They collectively form a one-hot vector that represents one of the three possible pen states:
\begin{enumerate}
\item $ p_1 = 1, \ p_2 = 0, \ p_3 = 0  \Rightarrow$ pen is currently touching a surface and line connecting previous point to this point will be drawn. 
\item $ p_1 = 0, \ p_2 = 1, \ p_3 = 0 \Rightarrow$ last point of a stroke, pen will be lifted  after this point.
\item $ p_1 = 0, \ p_2 = 0, \ p_3 = 1 \Rightarrow$ pen was lifted and stroke drawing ended.
\end{enumerate}
Neural networks are trained in batches for faster training, so it is important to describe how we mini-batch the strokes of drawings in our framework. We match each stroke of a symbol with strokes of other symbols according to their temporal order, then each of these matched sets comprise a single batch(i.e. first strokes are in one batch, second strokes are in another batch etc.). As an example, consider that we have three drawn objects: first one consisting of four strokes: $Symbol^1 = [stroke_1^1, \ stroke_2^1,$ $ \ stroke_3^1, \ stroke_4^1]$, the second sketch of two strokes: $Symbol^2 = [stroke_1^2, \ stroke_2^2]$, and a third having single stroke: $Symbol^3 = [stroke_1^3]$. Let our mini-batch size be equal to three (i.e. we have strokes of three sketches in a single batch), then these symbols will be batched into four batches each containing three strokes: 
\begin{itemize}
\item $batch_1 = [stroke_1^1; \ stroke_1^2; \ stroke_1^3]$
\item $batch_2 = [stroke_2^1; \ stroke_2^2; \ [0 \ 0 \ 0 \ 0 \ 1]'s]$
\item $batch_3 = [stroke_3^1; [0 \ 0 \ 0 \ 0 \ 1]'s; [0 \ 0 \ 0 \ 0 \ 1]'s]$
\item $batch_4 = [stroke_4^1; [0 \ 0 \ 0 \ 0 \ 1]'s; [0 \ 0 \ 0 \ 0 \ 1]'s]$ 
\end{itemize}
Points of strokes with same temporal order are put in the same batch. If some symbol has less strokes than other ones then $[0 \ 0 \ 0 \ 0 \ 1]'s$ are appended to a batch.
\subsection{Stroke-rnn}
\textsf{Stroke-rnn} is a sequence to sequence VAE similar to the one used by Ha and Eck. It consists of an \textit{encoder}  which produces a distribution over possible values of latent vector representations $z$ that could have been generated given stroke S (i.e. $\sim q(z|S)$), and a \textit{decoder} that given particular sampled vector $z$ generates a distribution over all possible corresponding values of S (i.e. $\sim p(S|z)$). Bi-RNN  was used for encoding and autoregressive RNN for decoding. Unlike Ha and Eck who used HyperLSTM \cite{ha2016hypernetworks} that is able to spontaneously augment its own weights, we use plain Long Short-Term Memory (LSTM) \cite{hochreiter1997long} network. Since our model is trained on strokes which are in general less complex and  shorter than complete symbols, the plain LSTM is more preferable than more complicated versions such as HyperLSTM. The \textsf{stroke-rnn} architecture is shown in Figure \ref{strokernn}.

Forward LSTM encodes stroke in normal sequential order and backward network encodes the stroke in a reverse order, we concatenate final hidden state vectors $h_f$ and $h_b$ to form single encoded vector $h$:
\begin{equation}
   h_f = encode_{f}(S) \quad  h_b = encode_{b}(S) \quad h = [h_f;\ h_b]
\end{equation}
Then we compute mean and standard deviation ($\mu \ and \ \sigma$) vectors of approximate posterior $\sim q(z|S)$, and sample a latent vector $z$ of size $N_z$ from the posterior: 
\begin{equation}
   \mu = W_\mu h + b_\mu, \ \hat{\sigma} = W_\sigma h + b_\sigma, \ \sigma = \exp( \frac{\hat{\sigma}}{2}), \  z = \mu + \sigma \odot \mathcal{N}(0, I)
\end{equation}
Where $\odot$ represents element-wise multiplication and $\mathcal{N}(0, I)$ is a zero mean unit variance Gaussian. We calculate the initial hidden state vector $h_0$ and the cell state vector $c_0$ of the decoder LSTM using latent vector $z$:
\begin{equation}
  [h_0; \ c_0] = \tanh(W_z z + b_z)
\end{equation}
In a decoding phase, at each time step $t \in 1, 2, .. L_s$ we concatenate the latent vector $z$ with a point vector $s_{t-1}$ and feed it to the decoder LSTM as an input. The starting point $s_0$ is defined as $[0 \ 0 \ 1 \ 0 \ 0]$. Then, we compute parameters of Gaussian Mixture Model(GMM) \cite{reynolds2015gaussian} that consists of $M$ Bivariate Gaussians and 3 pen states using decoder hidden state vector $h_t$ :   
\begin{equation}
y_t = W_y h_t + b_y
\end{equation}
where, $y_t$ is a $6M+3$ dimensional vector containing $M$ mixture weights, $2M$ ($i.e \ \mu_x \ \& \ \mu_y$) mean values, $2M$ ($i.e. \  \sigma_x \ \& \ \sigma_y$) standard deviations, $M$ correlation coefficients, and 3 pen states:
\begin{equation}
\begin{aligned}[b]
y_i = [(\hat{\Pi}_1 ... \hat{\Pi}_M) 
&(\mu_{x, 1} ... \mu_{x, M}) (\mu_{y, 1} ... \mu_{y, M}) 
(\hat{\sigma}_{x, 1} ... \hat{\sigma}_{x, M}) \\ 
&(\hat{\sigma}_{y, 1} ... \hat{\sigma}_{y, M})
(\hat{\rho}_{xy, 1} ... \hat{\rho}_{xy, M}) (\hat{q_1} \hat{q_2} \hat{q_3})]
\end{aligned}
\end{equation}
We exponentiate standard deviations in order to make them positive and squash correlation coefficients to have range between -1 and 1 using $\tanh$ function: 
\begin{equation}
\sigma_x = \exp(\hat{\sigma}_x), \quad \sigma_y = \exp(\hat{\sigma}_y), \quad \rho_{xy} = \tanh(\hat{\rho}_{xy})
\end{equation}
Next we normalize mixture weights $\Pi_k \ k \in 1, 2, ... M$ and pen state values $q_k \ k \in 1,2,3$ to have probabilities that sum up to 1 using softmax operation:
\begin{equation}
q_k = \frac{\exp(\hat{q}_k)}{\sum_{i=1}^{3}\exp(\hat{q}_i)} \quad \Pi_k = \frac{\hat{\Pi}_k}{\sum_{i=1}^{M} \hat{\Pi}_i}
\end{equation}
Now, we can sample next point from the GMM probability distribution:
\begin{equation}
p(\Delta x , \Delta y) = \sum_{i=1}^{M} \Pi_i \mathcal{N} (\Delta x , \Delta y)_i \quad where \quad \sum_{i=1}^{M} \Pi_i = 1
\end{equation}
The $\mathcal{N} (\Delta x , \Delta y)_i$ is a probability distribution function of the $i'$th Bivariate Gaussian in GMM defined as:
\begin{displaymath}
 \mathcal{N} (\Delta x , \Delta y )_i = \frac{1}{2\pi\sigma_{x,i}\sigma_{y,i}\sqrt[]{(1-\rho_{xy, i} ^ 2)}}\exp[-\frac{u_i}{2(1-\rho_{xy, i} ^ 2)}]
\end{displaymath}
\begin{displaymath}
and \quad  u_i = \frac{(\Delta x -\mu_{x, i})^2}{\sigma_{x,i}^2} + \frac{(\Delta y -\mu_{y,i})^2}{\sigma_{y,i}^2} - \frac{2\rho_{xy,i}(\Delta x -\mu_{x, i})(\Delta y -\mu_{y, i})}{\sigma_{x, i}\sigma_{y, i}}
\end{displaymath}
During training we feed actual stroke points as an input for RNN both in encoding and decoding phases. After training, we encode actual stroke points, then we reconstruct stroke by sampling points from probabilistic mixture model conditioned on latent vector $z$. At each decoding time step we concatenate point sampled at previous time step with the latent vector $z$ and feed the resulting vector as an input to the decoder. While sampling these points we can control the randomness of the GMM outputs using a temperature variable $\tau$:
\begin{equation}
\hat{q}_k \rightarrow \frac{\hat{q}_k}{\tau} \quad \hat{\Pi}_k \rightarrow \frac{\hat{\Pi}_k}{\tau} \quad \sigma_x^2 \rightarrow  \tau\sigma_x^2 \quad \sigma_y^2 \rightarrow \tau \sigma_y^2 
\end{equation}
As the temperature value approaches zero ($\tau \rightarrow 0$), the standard deviations of bivariate Gaussians also approach zero ($\sigma_x \rightarrow 0, \sigma_y \rightarrow 0$). In this mode the model becomes deterministic, the displacements $\Delta x \ \& \ \Delta y$ are sampled near mean $\mu_x \ \& \ \mu_y$ (i.e. most probable values) of the Gaussian with largest mixture weight $\hat{\Pi}_k$. On the other limit $\tau \rightarrow 1$ sampling process is completely random dependent on outputs of the decoder.

\subsection{Model training details}
Neural networks require well defined loss (or cost) functions for proper training. Our encoder-decoder network is trained end-to-end and it minimizes the sum of three different cost functions. The first cost function is a cumulative negative log-likelihood of displacement prediction probabilities over a point sequence of a stroke:
\begin{equation}
J_d = -\frac{1}{L_{s}} \sum_{t=1}^{L_s} \log(\sum_{i=1}^{M} \Pi_{i,t} \mathcal{N}(\Delta x_t, \Delta y_t )_i)
\end{equation}
The second loss is a cross entropy over the categorical probabilities of pen states:
\begin{equation}
J_{ps} = -\frac{1}{L_{s}} \sum_{t=1}^{L_{s}} \sum_{i=1}^{3} p_{i,t} \log(q_{i, t})
\end{equation}
Where $p_{t}$ is the correct one-hot pen state vector at time step $t$. The last optimization term is Kullback-Leibler (KL) divergence of estimated posterior from the true posterior: 
\begin{equation}
J_{KL} = -\frac{1}{2N_z} \sum_{i=1}^{N_z}(1 + \hat{\sigma}_i - \mu_i^2 - \exp(\hat{\sigma}_i))
\end{equation}
The overall loss function used in training is thus given by:
\begin{equation}
Loss = J_d + J_{ps} + w_{KL} J_{KL}
\end{equation}
Where $w_{KL}$ is the weight given to Kullback-Leibler divergence term. It is increased from starting $ w_{KLs}$ value to 1 by the following formula:
\begin{equation}
w_{KL} = 1 - (1 - w_{KLs})*R^{step}
\end{equation}
Multiplying $J_{KL}$ by small weight at the beginning of training procedure allows the optimizer to initially focus on reconstruction of correct displacements and pen states, and then focus on estimating posterior distribution. This strategy results in better overall losses \cite{sketchrnn}.

\subsection{Model setup}
We implemented the model using deep learning framework Knet \cite{knet2016mlsys}. We set training, validation and test set sizes to 70k, 2.5k and 2.5k respectively. We initialize the weights using Xavier                                    \cite{glorot2010understanding} initialization procedure.  To update parameters during training we use Adam \cite{kingma2014adam} optimizer with an initial learning rate of $10^{-4}$ and gradient clipping at 1.0. Also, we curtail overfitting by multiplying displacements $\Delta x $ and $\Delta y$ by a random number sampled uniformly from values between 0.9 and 1.1 and applying recurrent dropout without memory loss\cite{krueger2016zoneout} with keep probability of 90\%. For stable training dynamics and faster training time we utilize layer normalization \cite{ba2016layer} both for encoder and decoder recurrent neural networks. Moreover, unless otherwise stated we use the following settings - number of encoder hidden units: 512, decoder hidden state vector size: 1024, number of Gaussians in mixture model $M$: 20, latent vector size: $N_z$= 128, batch size: 100, starting KL weight $w_{KLs}$: 0.01 and KL divergence annealing coefficient $R$: 0.99995. 

\section{Experiments on Reconstruction Model} \label{Experiments on Reconstruction Model}
We train models on eight categories namely airplane, cat, chair, face, firetruck, flower, owl and pig. Then we perform two experiments to assess the reconstruction capabilities of the models.  
\subsection{Intra-category reconstruction}
\begin{figure}
\includegraphics[width=0.5\textwidth,height=\textheight,keepaspectratio]{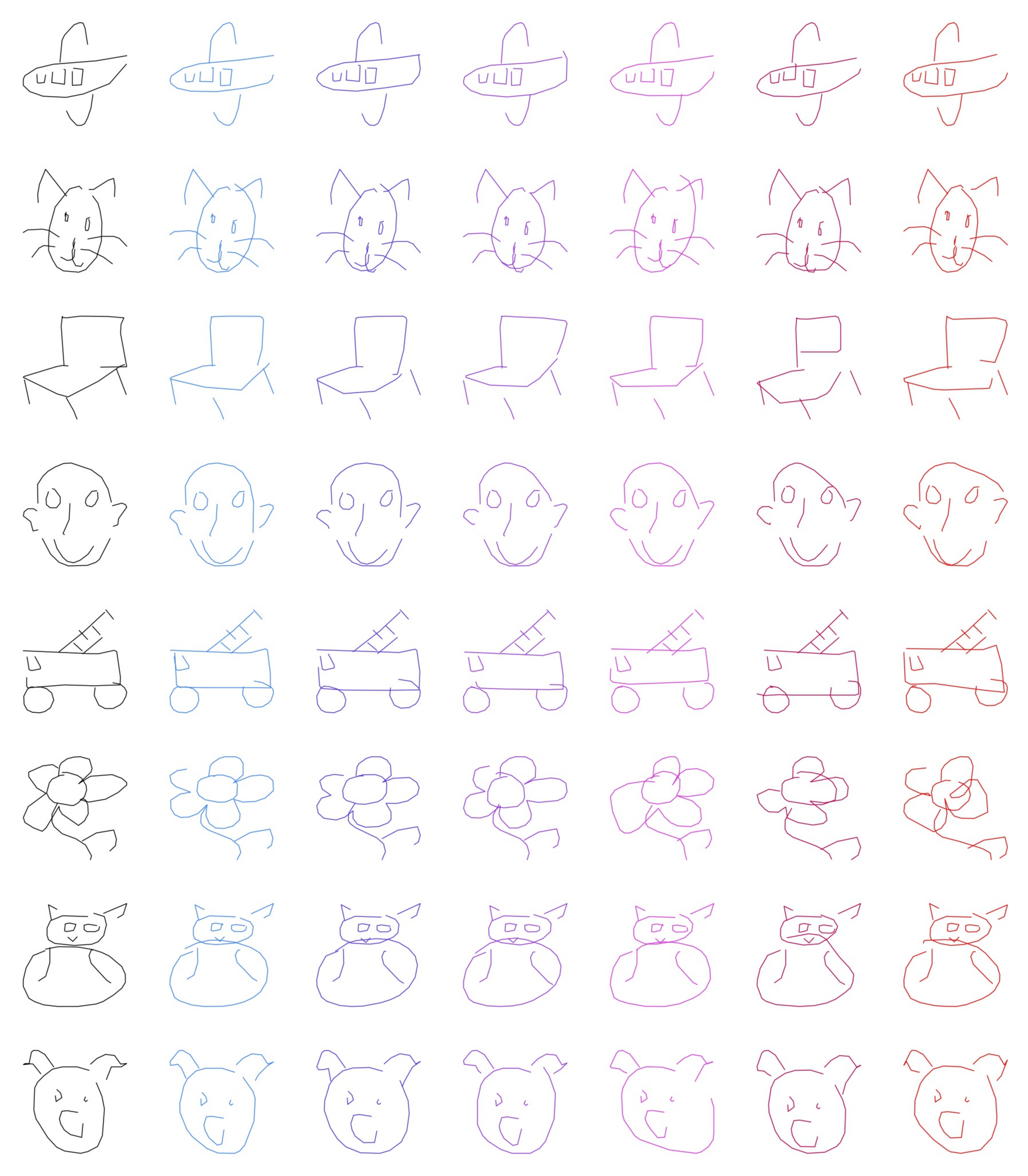}
\caption{Intra-category reconstructions with varying $\tau$ by the \textsf{stroke-rnn}.}
\label{intra-class}
\end{figure}

In this experiment we qualitatively assess reconstruction capabilities of trained models on their corresponding categories. Reconstruction results for various temperature values $\tau$ are shown in Figure \ref{intra-class}. The black drawings on left side represent actual human-drawn sketches. Model reconstructions as $\tau$ values linearly increase from 0.01 to 1 as we go from left (blue symbols) to right (red symbols) are displayed to the right of the original drawings.

In general, reconstructed symbols were pretty similar to real human inputs. Reconstructions for smaller values of $\tau$ tend to be more similar to actual symbols while for larger values of $\tau$ show some deviations form the input due to increased randomness of the model. In particular, flower decreased the number of petals from five to four as we increased $\tau$ from 0.01 to 1. Also, the chair model displayed different variations of the supporting back of the chair for bigger temperature values.  

\begin{figure}
\includegraphics[width=0.5\textwidth,height=\textheight,keepaspectratio]{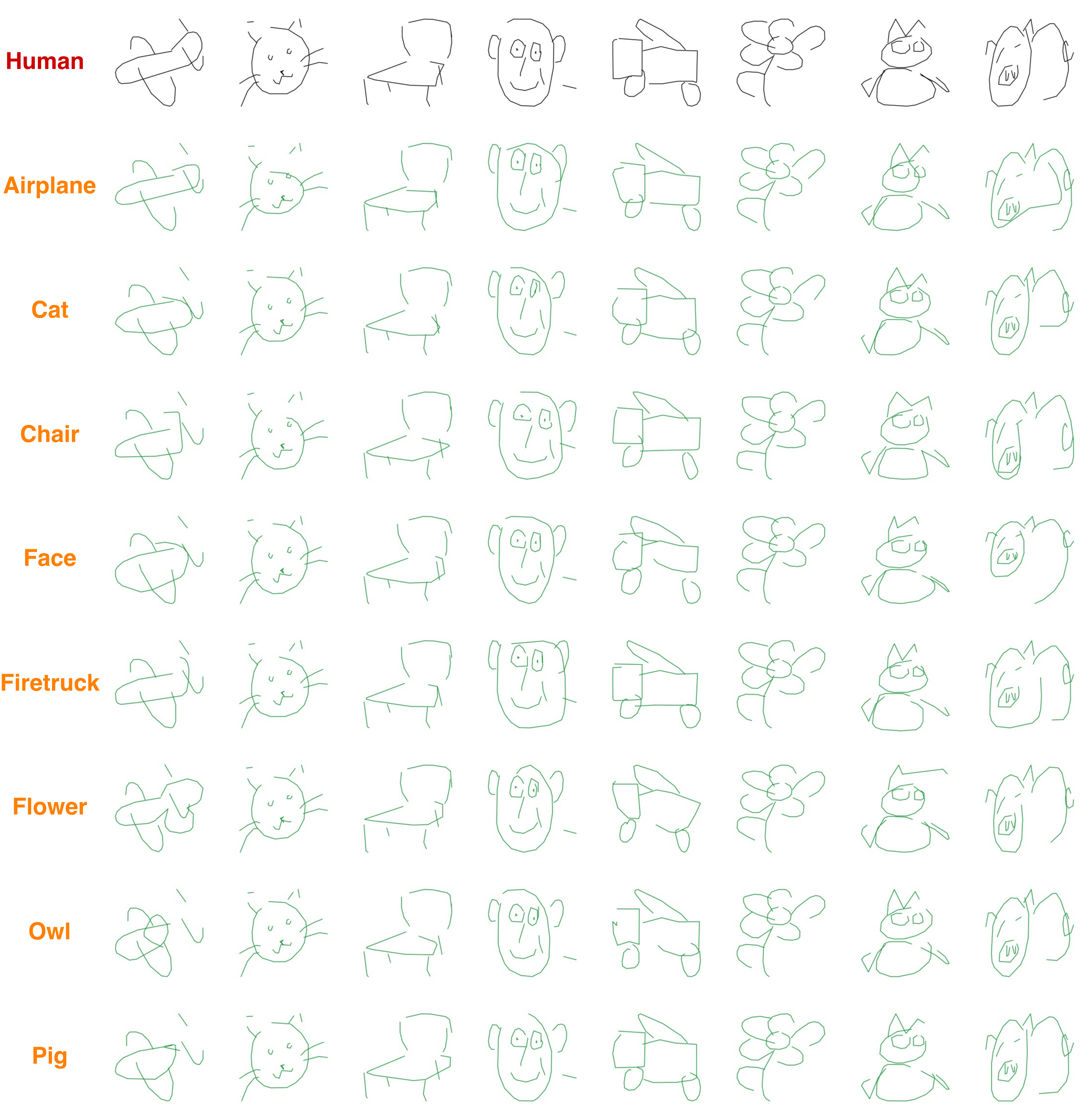}
\caption{Inter-category reconstructions by the \textsf{stroke-rnn}.}
\label{inter-class}
\end{figure}

\subsection{Inter-category reconstruction}
Real strength of our method is in it's capability of reconstructing multiple categories of symbols even if it is trained on a single category of drawings. In order to demonstrate this we perform cross-reconstruction experiment for trained models. In this experiment, each model generates symbols of all different categories including it's own. In Figure \ref{inter-class}, we demonstrate the results of this experiment. Temperature variable $\tau$ was set to 0.5 in order to make models neither too deterministic nor too random. Symbols on the first top row are actual human-drawn inputs, while drawings below them correspond reconstructions using various models. The model names are written on first leftmost column. 

Since our models are trained to learn lower level stroke representations instead of complete symbol embodiment, they have shown good reconstruction capabilities across various categories. As shown in Figure \ref{inter-class}, all models demonstrate good reconstruction capabilities on flower, cat and firetruck symbols which consist of most common strokes in forms of arcs, straight lines, circles, rectangles etc. Cat, pig and owl classes share many similar strokes (for ears, eyes, head etc.), so their models demonstrate almost perfect intra cross-reconstructions as well. Most of the models had trouble while reconstructing the peculiar stroke of the body of an airplane with a fuselage at its back. For instance, flower model has generated an airplane body that is hybrid of petals and body with fuselage. Hence, stroke-level framework could even be used to generate drawings with creative and peculiar strokes. In general, all of the models were able generate excellent human input reconstructions across various categories.

\section{Segmentation Model} \label{Segmentation Model}

\subsection{Datasets}
We evaluate our segmentation model on two different datasets. Although Huang at al. released their dataset of sketches with segmentation labels and we demonstrate success of our model on it, it was not suitable for our model since it contained only handful number of drawings per class, symbols were carefully drawn and they contained elaborate meaningless details. In contrast, symbols collected by Quick, Draw!, which were drawn in under 20 seconds, are simpler and realistic. Thus, in order to properly evaluate our proposed model, we annotated 500 sketches for 5 categories from Quick, Draw! dataset. We carefully selected sketches that are comprised of many semantically meaningful segments since vast number of drawings from QuickDraw! dataset lacked some components. This was due to the fact that users had time restrictions, also, most of them stopped to draw when the QuickDraw! background symbol recognition model was able to recognize their drawing. Moreover, we aimed to label set of symbols drawn in diverse styles. Therefore, while annotating, we inspected a lot of symbols to find the ones drawn in an uncommon style. Labeled symbol categories along with their segmentation classes are listed below:
\begin{itemize}
\item Airplane: body, tail, window, wing. 
\item Cat: body, ear, eye, head, leg, mouth, nose, tail, whisker. 
\item Chair: back, leg, seat.
\item Firetruck: body, cab, ladder, light, water hose, window, wheel.
\item Flower: core, leaves, petals, stem.
\end{itemize}

Annotated strokes represent either single semantically meaningful component or part of the component. However, we labeled few drawings containing a single stroke that spanned multiple components. For instance, because of the ubiquity of drawings that used single stroke that span both body and fuselage of an airplane we labeled strokes few of such drawings as body. Also, for the firetruck class many people drew cab and body with a single rectangularly shaped stroke rather than drawing cab explicitly. These strokes were labeled as body of the firetruck.
 
In order to compare with previous endeavors, we test our neural network model on a dataset collected by Huang et al. The dataset was used in evaluations by prior frameworks \cite{schneider2016example,huang2014data,LLi2018} and it contains 10 categories: chair, table, airplane, bicycle, fourleg, lamp, vase, human, candelabrum and rifle. For each category Huang et al. collected 30 symbols from 3 people (10 drawings per person); one of the users was an experienced artist and the other two were not. To collect a diverse dataset, drawers were provided reference images as an inspiration. These symbols are carefully drawn and they closely follow corresponding 3D meshes. Hence, they are much more complex drawings as compared to ordinary sketches that people usually draw.

Huang et al. provide many possible ground truth segmentations. Since they used pixel based method they justify this by the fact that some pixels belong to a stroke shared by several components and these pixels may have different interpretations. In experiments they compare their segmentation results to all possible ground truths and report best results. However, we randomly choose only a single ground truth since according to authors all possible ground truths are equally valid. Our results could only be made better if we reported best results among all possible ground truths.

\subsection{Our neural network model for symbol segmentation}

While drawing symbols humans often use strokes to represent semantically meaningful components. So, a human can be thought of as a generative model that can both generate and segment drawings. Since \textsf{stroke-rnn} can draw strokes in correct relative positions with appropriate shapes, it should also be able to encode necessary information for segmenting the sketch into meaningful components. In other words the representation learned while encoding can be utilized for segmentation purposes.

\begin{figure}
\includegraphics[width=0.5\textwidth,height=\textheight,keepaspectratio]{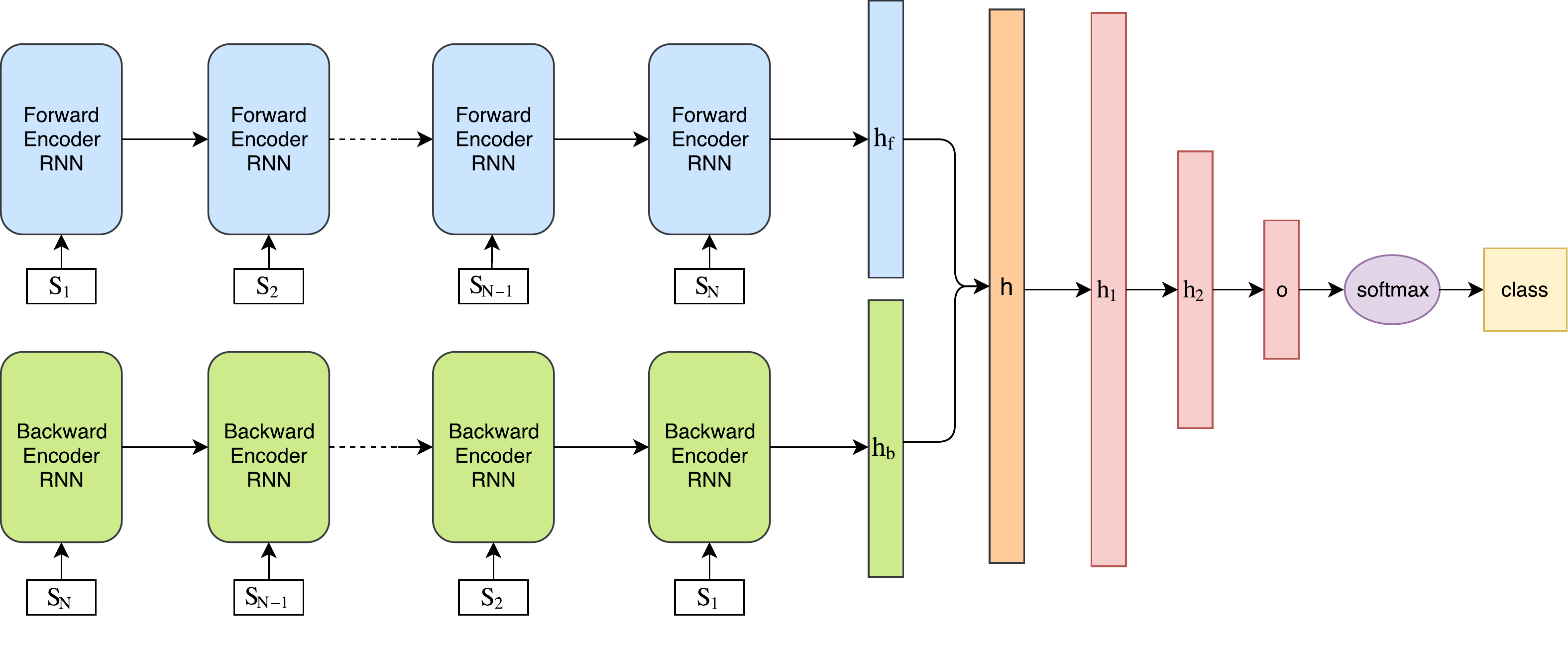}
\caption{Schematic diagram of our NN segmentation model.}
\label{segmodel}
\end{figure}

We reuse the encoder part of the \textsf{stroke-rnn} as a fixed (\textit{i.e. not updated during training process}) feature extractor and only train 3 layer MLP network as a segmentation model. As shown in Figure \ref{segmodel}, the encoder first generates the vector representation $h$ of the stroke, then $h$ is passed through two fully connected layers with dropouts after each layer:
\begin{equation}
h_1 = relu(W_{h_1} h + b_{h_1}), \quad h_2 = relu(W_{h_2} h_1 + b_{h_2})
\end{equation}
Finally, $h_2$ vector is passed through the last network layer and segmentation class probabilities are calculated using softmax:
\begin{equation}
o = W_{o} h_2 + b_{o},  \quad \hat{y}_c = \frac{\exp(o_c)}{\sum_{c=1}^{C}\exp(o_c)} \ for \ c \in {1, 2, .. C} 
\end{equation}
Where $C$ is number of number of segment classes for a particular symbol category and $\hat{y}_c$ is probability of class $c$ for a stroke.

\subsection{Baseline model}
We use an SVM with radial basis function (RBF) kernel  as a baseline model for evaluation of our dataset. We find best SVM model parameters using grid search over parameter space. 
\begin{figure*}
\includegraphics[width=\textwidth,height=\textheight,keepaspectratio]{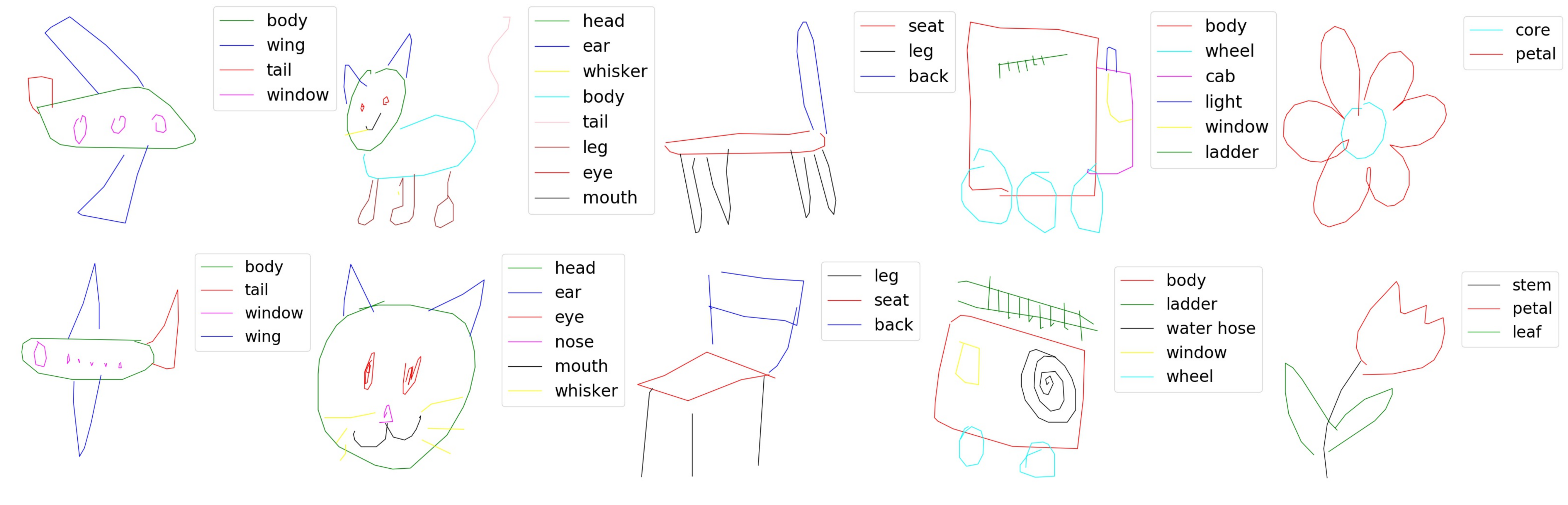}
\caption{Symbol segmentations by our NN framework.}
\end{figure*}
We employ Image Deformation Model (IDM) \cite{ouyang2009visual} feature vector for sketched symbol recognition. The vector contains 720 elements and was proven to be successful for many different settings of sketch recognition \cite{tirkaz2012sketched, ouyang2009learning, yesilbek2017sketch}. The feature is based on systematic mapping of the visual appearance of a drawing on five 2D matrices. There are four orientation feature matrices that correspond to four reference angles 0, 45, 90, and 135. They compute how nearly horizontal, vertical, or diagonal the stroke is, at each point. The feature value for each point varies linearly between 0 and 1 as the difference between orientation angle of current point and reference angle increases from 0 to 45, but if the angle difference is greater than 45 the value of 0 assigned to current point. Fifth feature assigns 1 to endpoint mappings of the stroke and 0 to all other points. These matrices are then linearized and concatenated to form final feature vector.

\subsection{ Model training details}
We keep feature extractor (encoder) fixed and train 3 layer MLP network. Since drawings have different number of strokes for each component class (e.g. generally head of a cat is drawn by single stroke and whiskers by six strokes), segmented dataset is imbalanced. Hence, we use weighted cross entropy loss that gives extra weight on learning correct segmentation of classes with smallest amount of instances. Let $n$ be a vector where $n_c$ is equal to number of strokes having label $c$, then we compute class weights $w_c$ as follows: 
\begin{equation}
 n = [n_{1}, n_{2}, .. n_{C}], \ w_c =  \frac{\exp(-\hat{n}_c)}{\sum_{c=1}^{C} \exp(-\hat{n}_c)} \ c \in 1,2,...C
\end{equation}
Where C is total number of classes for particular sketch. Then, weighted cross entropy loss function for segmentation model is:
\begin{equation}
J_{seg} = - \sum_{c=1}^C w_c y_c \log(\hat{y}_c)
\end{equation}
\subsection{Model setup}
First and second ($h_1 \& h_2 $) fully connected layer had 1024  and 512 nodes respectively. Number of nodes in output layer was equal to number of output classes C. We initialize the weights using Xavier initialization procedure. Unless otherwise stated we set the batch size to 16 and dropout probability to 50\%. Symbols used for training the \textsf{stroke-rnn} do not include component annotated symbols in any of the experiments.

\section{Experiments on Segmentation} \label{Experiments on Segmentation}
We present results of the segmentation experiments in this section.
\subsection{Segmentation}

We evaluate performance of neural network (our NN) model on the dataset that we labeled. As a baseline model for comparison we use plain IDM feature of each stroke and SVM with an RBF kernel for classification of strokes of a symbol. However, the spatial coordinates of the stroke are very helpful for segmentation of some categories. For example, the back of a chair is usually drawn above seat and legs are drawn below the seat. In order to make our feature stronger we concatenate coordinates of the start and end points of the stroke and mean of points of the stroke (i.e. center of mass) to our IDM feature vector (we call this new feature vector IDM+Spt). Yet, symbols can be drawn in various ways. For instance, a person asked to draw a cat may just draw a its head or sketch full embodiment with legs, tail, body, head etc. Hence, strokes appear at different locations depending on drawing style, and injecting context of the stroke in a symbol along with it's spatial location is advantageous for segmentation purposes. In order to capture this contextual information we concatenate the IDM of complete symbol, IDM of its stroke and aforementioned spatial coordinates to form a powerful feature representation (we call it IDM+Spt+Con). 
\begin{table}[!ht]
\begin{tabular}{cccccccc}
\hline
Category & IDM & IDM+Spt & IDM+Spt+Con & Our NN \\
\hline
Airplane    & 65.7 & 76.7 & 77.3 & 93.5  \\
Cat         & 68.6 & 74.4 & 78.5 & 85.8  \\
Chair       & 68.4 & 93.2 & 95.2 & 98.5   \\
Firetruck   & 74.3 & 78.0 & 81.3 & 86.4  \\
Flower      & 74.9 & 86.3 & 86.8 & 95.3 \\
 \hline
Average     & 70.4 & 81.7 & 83.8 & 91.9 \\
 \hline
\end{tabular}
\caption{\bf Segmentation accuracies on our dataset}
  \label{tab:Kurmanbek}
\end{table}

We compare performance of these SVM models to our NN model that utilizes encoder of the \textsf{stroke-rnn} which was trained on 70k symbols of corresponging category. To do so, we perform 5-fold cross validation for each model and the results are shown on Table \ref{tab:Kurmanbek}. The scores provided on the Table are component based accuracies, i.e. number of correctly labeled components divided by the total number of components. For all classes, as we expected, concatenating spatial coordinates to IDM feature significantly improved accuracies of SVM models. Addition of the whole symbol IDM as a context considerably increased accuracies for cat and firetruck as compared to other categories. This is due to the fact that cat and firetruck are more complex drawings than others since they could be drawn in more diverse styles. For instance, while petals of the flower are almost always drawn above the stem, the head of a cat might appear on either side or above the body of the cat. Therefore, addition of context of the stroke to a feature adds substantial information for segmentation purposes of these categories.

Overall, our NN framework outperforms all other models for all categories. On average, our NN model has 8.1\% higher accuracy than the best SVM-based model (i.e. with IDM+Spt+Con feature). Some of the segmentations by our neural network model are shown in Figure 5.
\subsection{Effect of annotated training set size}
\begin{figure}
\includegraphics[width=0.5\textwidth,height=\textheight,keepaspectratio]{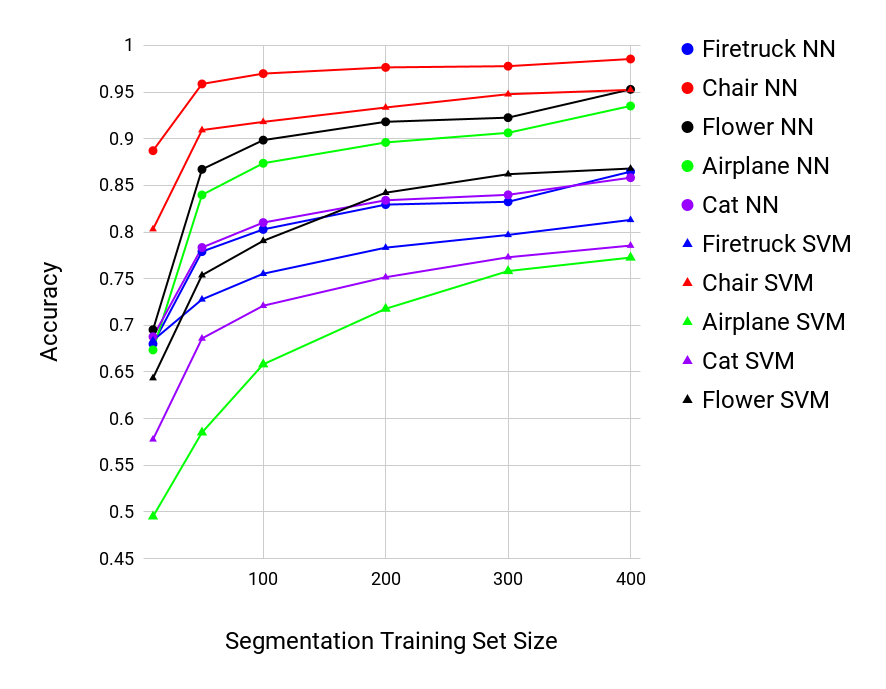}
\caption{Effect of training set size of annotated symbols on segmentation accuracy.}
\label{trnsetchange}
\end{figure}

In these experiments our goal is to identify the effect of training set size of symbols with component labels on performance. As previously we use encoders trained on 70k symbols of their corresponding categories as a fixed feature extractors. The training set sizes of component labeled symbols change as 10, 50, 100, 200, and 300; such that each subsequent bigger subset includes all symbols from preceding smaller subset. For each of those subsets we  compute average 5-fold cross validation accuracy and the results are shown in Figure \ref{trnsetchange}. We conduct the experiments for all categories in our dataset using best models based on SVM (i.e. with IDM+Spt.+Con feature) and our NN.  
In general, segmentation scores dramatically improve as the number of component annotated training samples increase from 10 to 100, and saturates after a subset size reaches 200 instances. This is expected since it becomes harder to capture details as variability of sketches increases. Note that our neural network model demonstrates higher accuracy than our best SVM-based model for all subset sizes. This also shows a merit to our approach since NN model is capable of performing well even on small training set sizes.

\subsection{Effect of training set size of the \textsf{stroke-rnn}}

\begin{figure}
\centering
\includegraphics[width=0.5\textwidth,height=\textheight,keepaspectratio]{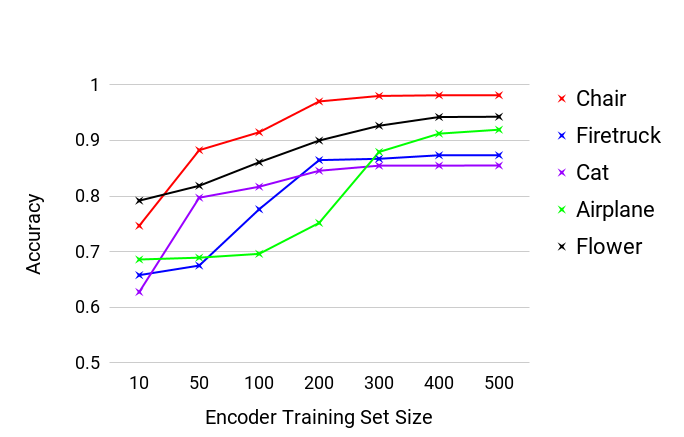}
\caption{Effect of training set size of the encoder on segmentation accuracy.}
\label{enctrnsetchange}
\end{figure}

In these experiments we assess importance of number of symbols that were used to train fixed feature extractor. To do so, we train the \textsf{stroke-rnn} with small number of training data and use the encoder of the model as a regular fixed feature extractor for segmentation experiments. In segmentation experiments we use all of the annotated symbols and perform 5-fold cross-validation. Segmentation accuracies are shown on Figure \ref{enctrnsetchange}. As our encoder training set size increases from 10 to 500 the segmentation accuracies increase and they all plateau at scores comparable to the scores of encoder trained on 70k sketched symbols (scores on Table \ref{tab:Kurmanbek}). Hence, the fixed encoder could be trained on much smaller amount (about 500 is enough) of unlabeled symbols to get good segmentation accuracies. 

\subsection{Effect of the encoder choice}

\begin{figure}
\centering
\includegraphics[width=0.5\textwidth,height=\textheight,keepaspectratio]{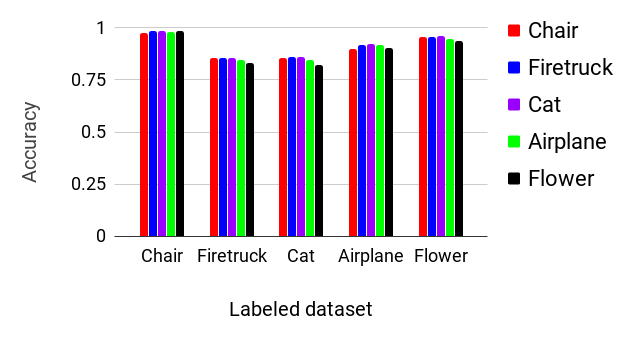}
\caption{Segmentation accuracies of encoders for each symbol category in our dataset.}
\label{crossmodelsegm}
\end{figure}

We study the performance of various encoders on different object categories of symbols. To do so, we take encoders trained on a single category of 500 unlabeled symbols and use them as a fixed feature extractor. Then, for each encoder we train a separate MLP segmentation network on each symbol category. Segmentation performances of different encoders (color-coded) on different object classes (horizontal axis) are reported on Figure \ref{crossmodelsegm}. Even if an encoder was trained on a single category (say cat) it can be reused as a fixed feature extractor for multiple other categories without significant impact on overall accuracy. This is expected since the encoder of the \textsf{stroke-rnn} is able to model multiple categories of symbols as we demonstrated previously.  

\subsection{Huang et al. dataset}

\begin{figure*}
\includegraphics[width=\textwidth,height=\textheight,keepaspectratio]{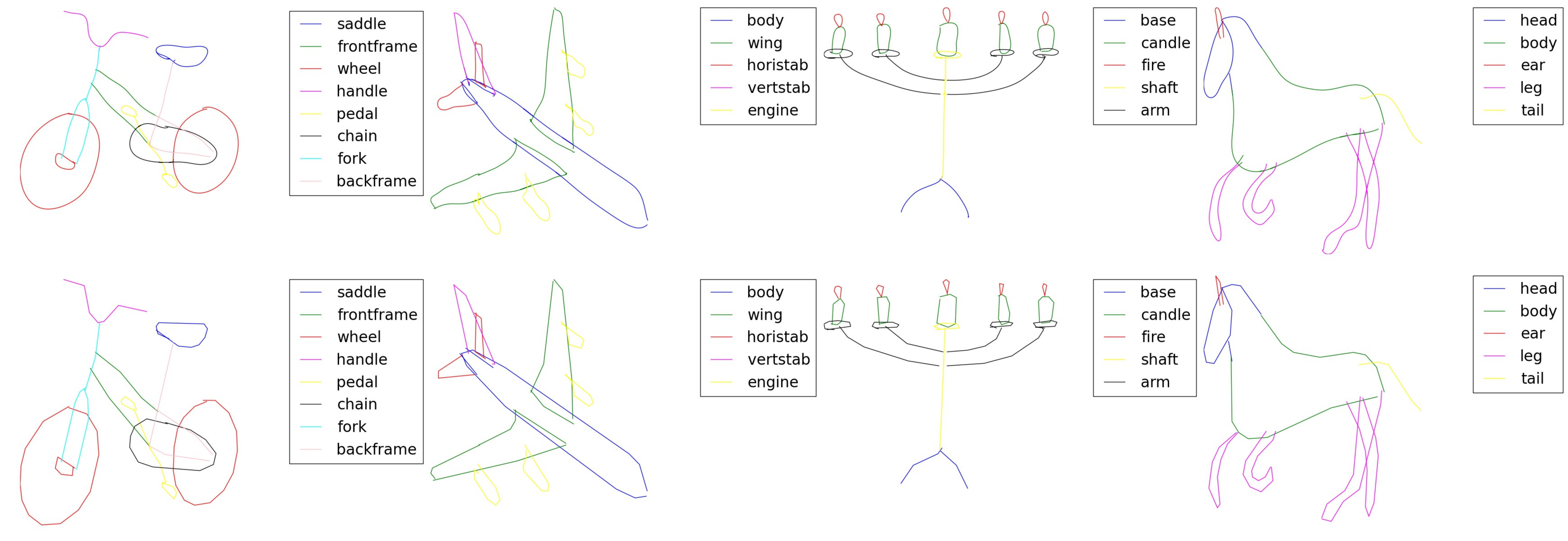}
\caption{Huang dataset preprocessing.}
\end{figure*}

To make drawings in the Huang et al. dataset compatible input for the stroke-rnn  we preprocess them in the same manner as Ha and Eck. In other words, we scale symbol points to have minimum value of 0 and maximum value of 255, resample all strokes with 1 pixel spacing and simplify all strokes using the Ramer-Douglas-Peucker \cite{douglas1973algorithms} algorithm with an epsilon value of 2.  Also, we remove semantically meaningless points or tiny straight lines with length smaller than 15 pixels. Some of the drawings before and after preprocessing are shown on upper and lower rows of Figure 6. The preprocessing procedure made stroke curves look less smooth but did not affect appearance of the sketches. We release to public the preprocessed data along with the labels that we use for evaluation.

Huang et al. segmented drawings by matching a sketch with a labeled mesh either manually (Huang) or automatically (Huang-A). They treated sketches as pixel images and considered a component to be correctly classified if 70\% of its pixels got appropriate label. Schneider and Tuytelaars used Fisher Vectors with Spatial coordinates and applied CRF to find the best global configuration of components (CRF). As a preprocessing step Schneider and Tuytelaars divided strokes into many smaller segments on high curvature points. Even though their components were not identical to the ones provided by Huang et al. we show their results too for completeness. Instead, we use the original component segmentation by Huang et al. and use 20 symbols for training and 10 for testing as in the original paper. We employ encoder of generative model of a firetruck as a fixed feature extractor and train separate fully connected layers for each category in the dataset. The segmentation accuracies for each category are demonstrated on Table \ref{tab:huangacc} with the best automatic results in boldface.
Average segmentation accuracies of our NN model were higher than reported scores by Huang et al., Schneider and Tuytelaars, and Li et al. for most of the categories. Average score across the classes was higher by about 5\% than even manually tuned (Huang) and CRF-based methods. Moreover, our model demonstrates about 2.4\% average improvement as compared to CNN-based model by Li et al..

\begin{table}[!ht]
\begin{tabular}{cccccccc}
\hline
Category & Huang & Hunag-A & CRF & Li et al. & Our NN\\
\hline
Airplane    & 66.2 & 55.8 & 48.7 & \textbf{76.9} & 67.7 \\
Bicycle     & 66.4 & 58.3 & 68.6 & \textbf{77.1} & 74.1 \\
Candelabra  & 56.7 & 47.1 & 66.2 & 69.9 & \textbf{78.4} \\
Chair       & 63.1 & 52.6 & 61.6 & \textbf{70.3} & 62.6 \\
Fourleg     & 67.2 & 64.4 & 74.2 & 75.5 & \textbf{81.1} \\
Human       & 64.0 & 47.2 & 63.1 & 72.8 & \textbf{84.3} \\
Lamp        & 89.3 & 77.6 & 77.2 & 83.8 & \textbf{90.3} \\
Rifle       & 62.2 & 51.5 & 65.1 & 65.7 & \textbf{66.7} \\
Table       & 69.0 & 56.7 & 65.6 & \textbf{77.3} & 73.2 \\
Vase        & 63.1 & 51.8 & 79.1 & 78.0 & \textbf{92.3} \\
 \hline
Average     & 66.7 & 55.3 & 67.0 & 74.7 & \textbf{77.1}  \\
 \hline
\end{tabular}
\caption{\bf Accuracies on the Huang et al. dataset. }
  \label{tab:huangacc}
\end{table}

\section{Summary and future work}

In this paper, first, we present a generative sketched symbol reconstruction model part of which we later reuse for segmentation purposes. Specifically, we proposed the VAE-based symbol drawing framework \textsf{stroke-rnn} that consists of Bi-RNN encoder and autoregressive RNN decoder. The decoder generates parameters for GMM conditioned on the latent vector $z$ which is sampled from the distribution produced by the encoder. The probabilistic nature of points sampled at each time step from the GMM distribution ensures that the model is capable draw strokes of various shapes. In relation to previous endeavors the \textsf{stroke-rnn} has two important advantages. First, it is trained to draw lower level stroke representation of sketches. Hence, it is able to generate multiple disparate categories of sketches even if it was trained on single category as shown in Figure 3. Moreover, our model is a natural choice for modeling higher quality drawings with large number of points. Symbols are generally composed of multiple strokes much shorter in length than a complete sketch, so it is better to train model on them since RNNs suffer from the problem of vanishing and/or exploding gradients \cite{bengio1994learning} when trained over long sequences. Training the \textsf{stroke-rnn} on high quality symbols possibly using more powerful RNN architectures like HyperLSTM is an interesting future direction.

Most importantly, this article explores stroke level symbol segmentation task which is very important research direction for semantic understanding of drawings. Since a good and comprehensive dataset of symbols with segmentation labels was not available, we annotated components for 5 categories from QuickDraw! dataset. During annotation process, we meticulously selected symbols in the dataset to represent diverse drawing styles. To our knowledge, we also proposed the first SVM-based model for segmentation of vector representation of sketches. We compared segmentation accuracies of the baseline SVM-based frameworks with different versions of the IDM feature to our NN model that is composed of encoder of a reconstruction model and 3 additional fully connected layers. The neural network model demonstrated 8.1 \% higher average accuracy compared to best baseline model on our dataset. In addition, our NN model achieves 2.4 \% average accuracy improvement over prior methods on Huang et al. dataset. Our NN model classified segments based solely on the feature extracted by the encoder. An interesting future extension to the segmentation model could be to encode the relationships between strokes and optimize the network to find globally optimal segmentation based on all strokes of the sketch. Moreover, another promising route would be to combine effective image processing ability of CNNs and sequence processing capability of RNNs to look for further improvements in segmentation accuracies.  

We investigated the effect of training set size of component labeled symbols on segmentation accuracy. The segmentation accuracies consistently increase for all categories as the training set size is increased from 10 to 200, and then the accuracy improvements begin to cease. One compelling future research direction could be to study possible methods to increase accuracies for training set sizes of beyond 200 sketches. On the opposite side, another interesting study could be done to increase segmentation accuracies for small training set sizes. A good segmentation method working with few training examples would alleviate the burden of annotation that researchers often face. In experiments with fewer training samples we randomly select symbols to use for training. While, Yanik and Sezgin \cite{yanik2015active} show that a strategic selection of more informative drawings via guidelines from Active Learning literature improves sketch classification accuracies. An application of Active Learning selection protocols would potentially boost performance of sketch segmentation for few training instances. Also, we studied the effect of the training set size of the generative model's encoder on the segmentation performance. Best accuracies were achieved even for a small size of 500 symbols. Moreover, since the \textsf{stroke-rnn} learns lower level stroke representations and can draw multiple categories, an encoder trained on a single category could be used as a fixed feature extractor for different segmentation networks with negligible effects on accuracy as demonstrated on Figure \ref{crossmodelsegm}. Hence, overall training time of the segmentation network could be significantly reduced by training a single encoder on relatively small number of symbols and reusing it as fixed feature extractor for MLP segmentation networks of different categories.

We believe that variety of sketch-based applications could benefit from our results. The \textsf{stroke-rnn} trained on higher quality drawing could be used for educational purposes as a step by step sketching guide. The segmentation model could complement the sketching guide as means for providing meaningful feedback for drawing each component in a correct manner. We also envisage that part level segmentation of sketches would benefit sketch-based image and video retrieval systems. Combining sketch segmentation and image segmentation methods would make it possible to accurately map the sketch to candidate images or frames. On a different setting part level labels could be used to generate text queries for search systems. Moreover, adding automatic labeling could help to create fluid animations. For instance, it would help an effective morphing of related components during animation. In summary, we believe that there are many potential applications that could benefit from this work.

%
%
%
%

\bibliographystyle{ACM-Reference-Format}
\bibliography{body-bibliog.bib}
\end{document}